\documentclass[10pt,a4paper,sort&compres]{article}

\usepackage{authblk}
\usepackage{graphicx}
\usepackage[margin=1in]{geometry} % full-width

\usepackage{amssymb}
\usepackage[colorlinks=true,linkcolor=blue,citecolor=blue,urlcolor=blue]{hyperref}
\usepackage[table,xcdraw]{xcolor}

\title{Exploring the nonclassical dynamics of the\\ ``classical'' Schr\"odinger equation}
		
\author{David Navia$^*$ and \'Angel S. Sanz$^\dagger$}

\affil{Department of Optics, Faculty of Physical Sciences, Universidad Complutense de Madrid\\ Pza.\ Ciencias 1, 28040 Madrid, Spain\\
e-mail: $^*$junavia@ucm.es, $^\dagger$a.s.sanz@fis.ucm.es}

\date{}

\begin{document}

\maketitle
		
		\begin{abstract}
			The introduction of nonlinearities in the Schr\"odinger equation has been considered in the
			literature as an effective manner to describe the action of external environments or mean fields.
			Here, in particular, we explore the nonlinear effects induced by subtracting a term proportional to
			Bohm's quantum potential to the usual (linear) Schr\"odinger equation, which generates the so-called
			``classical'' Schr\"odinger equation.
			Although a simple nonlinear transformation allows us to recover the well-known classical
			Hamilton-Jacobi equation, by combining a series of analytical results (in the limiting cases) and
			simulations (whenever the analytical treatment is unaffordable), we find an analytical explanation
			to why the dynamics in the nonlinear ``classical'' regime is still strongly nonclassical.
			This is even more evident by establishing a one-to-one comparison between the Bohmian trajectories
			associated with the corresponding wave function and the classical trajectories that one should
			obtain.
			Based on these observations, it is clear that the transition to a fully classical regime requires
			extra conditions in order to remove any trace of coherence, which is the truly distinctive trait
			of quantum mechanics.
			This behavior is investigated in three paradigmatic cases, namely, the dispersion of a free
			propagating localized particle, the harmonic oscillator, and a simplified version of Young's
			two-slit experiment.
		\end{abstract}
		
%		\begin{keyword}

			
			
			
%		\end{keyword}
		
%	\end{frontmatter}
	
	

\date{}

	%%%%%%%%%%%%%%%%%%%%%%%%%%%%%%%%%%%%%%%%%%%%%%%%%%%%%%%%%%%%%%%%%%%%%%%%
	%%%%%%%%%%%%%%%%%%%%%%%%%%%%%%%%%%%%%%%%%%%%%%%%%%%%%%%%%%%%%%%%%%%%%%%%

\section{Introduction}
\label{sec1}

Typically, the emergence of the classical world from the quantum one is associated with
the action of decoherence
\cite{zurek:PhysToday-rev:2003,giulini-bk,schlosshauer:RMP:2004,schlosshauer:PhysRep:2019,schlosshauer-bk:2007}.
Accordingly, the constant interaction with an environment is what eventually leads a
given (quantum) system of interest to loose its quantumness (its ability to display
genuine quantum features, such as interference, for instance), and hence to behave
like the objects that we observe in our everyday life.
The theory of open quantum systems \cite{breuer-bk:2002} contains a wide variety
of effective models to investigate the effects and consequences arising from the
environmental interaction, while avoiding the technical complexities (analytical
and numerical) inherent to the corresponding many-body problems.
Among all such models, the Lindblad equation provides us with the most general form
of a master equation, where the environmental action on the system reduced density
matrix is described by a dissipator term.
Nonetheless, if we are interested in a wave-function based description, still the
Lindblad equation can be recast in terms of a stochastic nonlinear equation, according
to the quantum state diffusion approach \cite{diosi:JPA:1988,gisin:JPA:1992,percival-bk}.

To investigate the quantum-to-classical transition, however, there is an alternative
route, which consists in considering effective nonlinear equations.
This is the case of the so-called ``classical'' Schr\"o\-din\-ger equation,
\begin{equation}
	i\hbar \frac{\partial \psi ({\bf r},t)}{\partial t} =
	- \frac{\hbar^2}{2m} \nabla^2 \psi ({\bf r},t)
	+ V({\bf r}) \psi({\bf r},t)
	+ \frac{\hbar^2}{2m} \frac{\nabla^2 A({\bf r},t)}{A({\bf r},t)} \psi({\bf r},t) ,
	\label{eq5}
\end{equation}
where $A({\bf r},t)$ denotes the amplitude of the wave function $\psi({\bf r},t)$.
In this equation, the effect of introducing a nonlinear term on the right-hand side gives rise
to a phase dynamics governed by a classical-like Hamilton-Jacobi equation.
It is worth noting that the introduction of nonlinear terms
into the Schr\"odinger equation in this manner has been considered in the literature as an effective method to recover or reproduce classical-like behaviors.
For instance, the Kostin or Schr\"odinger--Langevin equation
\cite{kostin:jcp:1972,katz:AOP:2016,schleich:JLowTempPhys:2023}, or
the Schuch--Chung--Hartmann equation
\cite{schuch:JMathPhys:1983,schuch:IJQC:1984-1,schuch:JMathPhys:1984-2} constitute
attempts to describe dissipation in a pure quantum state, in an analogous manner
to how it is done in classical mechanics, with a friction term.
In the first case, in particular, the dissipative term is associated with the phase
of the wave function, while in the latter it is related to the square of its
amplitude (i.e., the probability density).
The so-called Schr\"odinger-Newton equation is another example of effective nonlinear
equation, formerly intended to investigate the equilibrium configurations of
self-gravitating systems of scalar bosons and spin$-1/2$ fermions, and later reconsidered
to study the role of gravity in the collapse of the wave function (the so-called problem
of the quantum state reduction) and, therefore, how classicality emerges in a cosmological
scenario \cite{diosi:PLA:1984,penrose:GenRelGrav:1996}.
A simplification of this equation is the Gross--Pitaevskii equation \cite{pethick2008bose}, of common used to study the dynamics of Bose-Einstein condensates, although it can also be extended to other contexts, such as the propagation of light through a nonlinear medium \cite{michinel:PhysicaD:2020}.
In this equation, the self-interacting term becomes a cubic nonlinearity, which acts on the system wave function only locally, that is, it is the density at a given position and time what acts on the wave function at such a position and time.
This equation is also used to described light propagation along optical fibers, when it is derived from Maxwell's equations \cite{Biswas2010-ch,fibich-bk}, instead of from a collection of bosons.
Nonetheless, in this context, nonlinearities can be just a cubic term, as in the Gross--Pitaevskii equation, or acquire a more generalized and complicated form in order to include other effects, such as ultrashort pulses or multiphoton absorption, which lead to the appearance of higher orders in the time-derivatives or imaginary fractionary powers of the nonlinear term.

Following the Van Vleck semiclassical formulation of quantum mechanics \cite{vanVleck:PNAS:1928},
Schiller found \cite{schiller:PhysRev:1962} that by modifying the classical Hamilton-Jacobi
equation one can obtain a classical Schr\"odinger equation with classical complex-valued wave
functions.
Almost by the same time, Rosen also found the same equation while investigating when quantum
and classical dynamics look the same
\cite{rosen:AJP-1:1964,rosen:AJP-2:1964,rosen:AJP:1965,rosen:FoundPhys:1986}.
A heuristic derivation of this equation employing the language of Lagrangian dynamics is given
by Holland \cite{holland-bk}, who also discusses a series of aspects related to it and its
solutions concerning the dissimilarities between quantum and classical motions (``classical''
in terms of this nonlinear equation).
On the other hand, following similar arguments to those of Rosen, Ghose \cite{ghose:FoundPhys:2002} reconsiders the classical Schr\"odinger equation as a means to recover classicality alternative to decoherence or other classical approaches, such as $\hbar$ going to zero.
More recently, it has also been revisited by Schleich {\it et al.}~\cite{Schleich:PNAS:2013},
from whom ``the linearity of quantum mechanics is intimately connected to the strong coupling
between the amplitude and phase of a quantum wave'', which is, actually, what the Bohm's picture
of quantum mechanics emphasizes.
On the other hand, numerical simulations of well-known paradigmatic quantum systems carried out
by Chou \cite{chou:AOP:2016,chou:AOP:2018}, Benseny {\it et al.}~\cite{oriols:FNL:2016}, and Ghose and Bloh \cite{ghose:arxiv:2017} show
that, despite of the strong connection between the classical Schr\"odinger equation and the
classical Hamilton-Jacobi, there are still major discrepancies between the motions generated
by one and the other.
On the other hand, Richardson {\it et al.}~\cite{richardson:PRA:2014} showed that still it is possible to keep linear features by conveniently tuning the contribution of the nonlinear term in the classical Schr\"odinger equation.

To better understand what really makes interesting the classical Schr\"odinger equation, it is
worth noting the fact that it includes a locally evaluated mean or self-consistent field term
that antagonizes the dispersive contribution implicit in the kinetic operator.
Thus, like Schiller \cite{schiller:PhysRev:1962}, let us consider the system wave function
in polar form,
\begin{equation}
 \psi({\bf r},t) = A({\bf r},t) e^{i S({\bf r},t)/\hbar} ,
 \label{eq1}
\end{equation}
where both $A({\bf r},t) = \sqrt{\rho ({\bf r},t)}$ and $S({\bf r},t)$ are real-valued fields
describing, respectively, the local variations of the probability amplitude (or the probability
density $\rho$) and the phase undergone by the system.
The action of the kinetic operator, $\hat{\mathcal{K}} = - (\hbar^2/2m) \nabla^2$, on this ansatz,
divided by $\psi ({\bf r},t)$, produces
\begin{equation}
 \frac{\hat{\mathcal{K}} \psi({\bf r},t)}{\psi ({\bf r},t)} =
 \frac{[\nabla S({\bf r},t)]^2}{2m} - \frac{\hbar^2}{2m} \frac{\nabla^2 A({\bf r},t)}{A({\bf r},t)}
	- \frac{i\hbar}{2} \frac{\nabla {\bf j}({\bf r},t)}{\rho ({\bf r},t)}\ .
 \label{eq2}
\end{equation}
On the right-hand side of the above equation, we have
\begin{equation}
	{\bf j}({\bf r},t) = \rho ({\bf r},t) {\bf v}({\bf r},t)
	= \frac{1}{m}\ {\rm Re} \left[ \psi^*({\bf r},t) \hat{\bf p} \psi({\bf r},t) \right] ,
\end{equation}
which is the usual quantum flux \cite{schiff-bk}, where $\hat{\bf p} = - i\hbar\nabla$ is the usual
momentum operator, and
\begin{equation}
	{\bf v}({\bf r},t) = \frac{{\bf j}({\bf r},t)}{\rho ({\bf r},t)}
	= \frac{1}{m}\ {\rm Im} \left[ \frac{\hat{\bf p} \psi({\bf r},t)}{\psi({\bf r},t)} \right]
	= \frac{\nabla S({\bf r},t)}{m}
 \label{eq4}
\end{equation}
is a velocity field that accounts for the local phase variations.
Within the context of Bohmian mechanics, this velocity field is directly related to the so-called
Bohm's momentum \cite{bohm:PR:1952-1}.

From a physical point of view, the three terms in Eq.~(\ref{eq2}) contribute to a different
dynamical aspect displayed by the wave function during its evolution \cite{sanz:JPA:2008}.
The imaginary term, which depends on the quantum flux and the probability density, is related to
the conservation of the latter.
As for the two real-valued contributions, they are specifically related to the wave function
propagation dynamics.
In analogy to the role of the kinetic term within the classical Hamilton-Jacobi formulation, the
first one of these contributions is also a purely kinetic term, i.e., it is connected to the ``motion''
of the wave function in the corresponding configuration space.
However, because a wave function is an extensive object, its dynamics is also strongly influenced by
its own configuration, which is precisely what the second term accounts for, as it depends on the
local changes undergone by the amplitude of the wave function (more specifically, it is a measure
of its local curvature at a given time $t$).
In the literature, this term is the so-called Bohm's quantum potential
\cite{bohm:PR:1952-1,bohm-hiley-bk,holland-bk}, although the term ``potential'' might be certainly
misleading, as neither its origin nor its action on the system have to do with those of a usual
potential function.
Note that it is responsible for dispersion or diffraction of the wave function, thus governing
its spreading \cite{sanz:JPA:2008,sanz-bk-2}.

The idea of a classical Schr\"odinger equation thus departs from the above considerations: if an equivalent term, but with opposite sign, is directly added to the usual Schr\"odinger equation, then it should counteract the dispersive effects induced by the quantum potential generated by the kinetic operator and, therefore, the dynamics displayed by the quantum system should look pretty much like classical
dynamics, since only the classical-like contribution, $(\nabla S)^2/2m$, remains.
Such a term is what Schleich {\it et al.}~\cite{Schleich:PNAS:2013} call the classicality-enforcing potential, a term also adopted by Richardson {\it et al.}~\cite{richardson:PRA:2014}.
The purpose in this work is to investigate this conjecture, which, in principle, seems to solve the
problem of the classical limit and the quantum-classical correspondence, as trajectories can also be
introduced in quantum mechanics through Eq.~(\ref{eq4}).
Specifically, by integrating in time the equation of motion $\dot{\bf r} = {\bf v}({\bf r},t)$, we
obtain the so-called Bohmian trajectories in a natural manner \cite{sanz:FrontPhys:2019}, which here
are interpreted as paths or streamlines that serve to monitor the flux of the probability density
\cite{sanz:AJP:2012} rather than the actual paths followed by real particles.
Note that, if the polar ansatz (\ref{eq1}) is substituted into the nonlinear Schr\"odinger equation
(\ref{eq5}), we obtain
\begin{eqnarray}
 - \frac{\partial S({\bf r},t)}{\partial t} & = & \frac{[\nabla S({\bf r},t)]^2}{2m} + V({\bf r}) ,
 \label{eq6} \\
 \frac{\partial \rho ({\bf r},t)}{\partial t} & = & - \nabla {\bf j}({\bf r},t) ,
 \label{eq7}
\end{eqnarray}
where Eq.~(\ref{eq7}) is the continuity equation, while Eq.~(\ref{eq6}) looks like a classical
Hamilton-Jacobi equation.

Unlike previous works \cite{chou:AOP:2016,chou:AOP:2018,oriols:FNL:2016}, here we are interested in
providing analytical solutions to also paradigmatic cases in order to better understand the main
differences between the quantum (Bohmian) solutions and the classical ones, and, more specifically,
such differences come from.
In particular, the analysis will focus on some paradigmatic systems, such as dispersion of a free
particle, the harmonic oscillator, or two-wave packet interference, as all of them capture in one
way or another the essence of what we understand by non-classical behavior in a simple manner, even
though they are often invoked to explain more complex problems.
To tackle the issue, we revisit the transition from a fully quantum or linear regime, described by
the usual Schr\"odinger equation, to the (seemingly) classical one determined by the nonlinear
Eq.~(\ref{eq5}) by adding a coupling factor, $\lambda$, to the nonlinearity.
This coupling factor ranges from 0 to 1, which allows us to pass from the linear to the nonlinear case
in a smooth manner.
To proceed in an analytical manner in as much as possible, at least, in the limiting cases, we have
considered Heller's frozen Gaussian wave-packet method \cite{heller:JCP:1975}.
According to this method, the propagation of a Gaussian wave packet can be determined by means of a
a series of time-dependent parameters that obey a set of coupled ordinary differential equations,
to which the Bohmian equation of motion can also be added \cite{sanz-bk-2}, which acquires a rather
simple form.
In some particular instances, these equations admit analytical solutions, which makes the interpretation
of the corresponding dynamics more evident, even in those cases where the wave function consists of
coherent superpositions of such wave packets and the full solution is no longer analytical.
Thus, the results obtained show that, although the overall classical behaviors are nicely reproduced,
there are still major differences that avoid us to speak about a true classical limit, in agreement
with previously reported data.
In this regard, even though a classical-like equation is recovered from the nonlinear Schr\"odinger
equation (\ref{eq5}), the dynamics still keeps strong non-classical features.

The organization of this work is as follows.
In next section, we briefly introduced the equations of motion in the general case.
The dynamics for single wave packets is discussed in Sec.~\ref{sec3}, considering both free
propagation and the harmonic oscillator.
In Sec.~\ref{sec4} we present results for two wave-packet interference propagating both in free
space and inside a harmonic potential.
Finally, the main conclusions extracted from this work are presented in Sec.~\ref{sec5}.

%%%%%%%%%%%%%%%%%%%%%%%%%%%%%%%%%%%%%%%%%%%%%%%%%%%%%%%%%%%%%%%%%%%%%%%%
%%%%%%%%%%%%%%%%%%%%%%%%%%%%%%%%%%%%%%%%%%%%%%%%%%%%%%%%%%%%%%%%%%%%%%%%

\section{General framework}
\label{sec2}

For simplicity, let us consider the case of a one-dimensional non-relativistic particle with mass $m$
acted by the nonlinear Schr\"odinger equation
\begin{equation}
 i\hbar \frac{\partial \psi (x,t)}{\partial t} =
 - \frac{\hbar^2}{2m} \frac{\partial^2 \psi (x,t)}{\partial t^2} + V(x) \psi(x,t)
	+ \lambda\ \frac{\hbar^2}{2m} \frac{1}{A(x,t)} \frac{\partial^2 A(x,t)}{\partial x^2}\ \psi(x,t) ,
	\label{eq8}
\end{equation}
where the potential function $V(x)$ can be recast as a second-degree polynomial (higher order
polynomials are disregarded here because of their loss of analyticity, as discussed below) and
the parameter $\lambda$ determines the strength of the coupling with the nonlinear contribution,
with its value ranging between 0 (linear regime) and 1 (classical regime).
Analogous gradual transitions can also be found in the literature \cite{chou:AOP:2016,chou:AOP:2018,ghose:arxiv:2017,richardson:PRA:2014}.
Regardless of the nonlinear term, the equation of motion for the trajectories describing the
time evolution of systems governed by Eq.~(\ref{eq8}) is
\begin{equation}
 \dot{x}(x,t) = \frac{j(x,t)}{\rho(x,t)} = \frac{1}{m} \frac{\partial S(x,t)}{\partial x} .
 \label{eq9}
\end{equation}
Given the degree of the potential, let us then consider the suitable Gaussian ansatz
\begin{equation}
 \psi (x,t) = \exp \left[ \frac{i\alpha_t  (x - x_t)^2}{\hbar} + \frac{i\beta_t (x - x_t)}{\hbar} + \frac{i\gamma_t}{\hbar} \right] .
 \label{eq10}
\end{equation}
which depends on the parameters $\alpha_t$, $\beta_t$, $\gamma_t$, and $x_t$.
The justification for this ansatz arises \cite{heller:JCP:1975} from the fact that, if the
wave packet is sufficiently narrow for a time and it is acted by a relatively smooth external
potential, then the dynamics of such a wave packet will only be affected (for such a time
duration) by the first terms in the Taylor expansion of the potential around its centroidal
position, according to Ehrenfest's theorem \cite{schiff-bk}.
That is, if we retain terms up to $(x - x_t)^2$ in this Taylor expansion, then the wave packet
will be acted, in good approximation, by a harmonic potential, which means that initially
Gaussian wave packets will preserve their Gaussian shape.

The ansatz (\ref{eq10}) allows us to obtain the solutions to the nonlinear Schr\"odinger equation
(\ref{eq8}) from a set of partial differential equations instead of solving directly a nonlinear
partial differential equation.
Those equations arise after substituting (\ref{eq10}) into Eq.~(\ref{eq8}), with the corresponding
approximation, up to second order, for the potential function around $x_t$, i.e.,
\begin{equation}
 V(x,t) = V_t (x_t) + V'_t (x_t) (x - x_t) + \frac{1}{2}\ V''_t (x_t) (x - x_t) ,
 \label{11a}
\end{equation}
and then separating in terms of the powers of $(x-x_t)$.
Moreover, let us also consider a series of conditions that simplify the calculations.
From the normalization condition for the ansatz (\ref{eq10}), we obtain a relationship
between the imaginary parts of $\alpha_t$ and $\gamma_t$:
\begin{equation}
 \gamma^i_t = - \frac{\hbar}{4}\
   \ln \left[ \frac{2 \alpha^i_t}{\pi\hbar} \right] ,
 \label{eq11}
\end{equation}
where a superscript ``i'' is used to denote imaginary part (similarly, a superscript ``r'' will be used
to label the real part of a quantity from now on).
Since the condition (\ref{eq11}) must be satisfied at any time and, at $t=0$ it implies the initial condition
\begin{equation}
	\gamma^i_0 = - \frac{\hbar}{4}\
	\ln \left[ \frac{2 \alpha^i_0}{\pi\hbar} \right] .
	\label{eq12}
\end{equation}
On the other hand, taking into account (\ref{eq11}), if we compute the expectation value for the position
and momentum of (\ref{eq8}), we obtain
\begin{equation}
 \langle \hat{x} \rangle = x_t , \qquad \langle \hat{p} \rangle = \beta_t .
 \label{eq14}
\end{equation}
Accordingly, both $x_t$ and $\beta_t$ are real-valued quantities, which determine the instantaneous
position and momentum of the centroid of the wave packet centroid in phase space.

Proceeding now as it was indicated above, we obtain, respectively, from the zeroth, first, and second
orders in $(x - x_t)$ the following equations
\begin{eqnarray}
 \dot{\gamma}_t & = & \frac{i\hbar\alpha_t}{m} + \beta_t \dot{x}_t - \frac{\beta_t^2}{2m} + V_t
  + \frac{\lambda\hbar}{m}\ \alpha^i_t
  - \frac{\lambda}{2m} \left( \beta^i_t \right)^2 ,
  \label{eq15} \\
 2\alpha_t \left( \dot{x}_t - \frac{\beta_t}{m} \right) & = & \dot{\beta}_t + V'_t
  + \frac{2\lambda\hbar}{m}\ \alpha^i_t \beta^i_t ,
  \label{eq16} \\
  \dot{\alpha}_t & = & - \frac{2\alpha_t^2}{m} - \frac{1}{2}\ V''_t
  - \frac{2\lambda}{m} \left( \alpha^i_t \right)^2 .
 \label{eq17}
\end{eqnarray}
Because $\beta_t$ is real, according to (\ref{eq14}), Eq.~(\ref{eq16}) reduces to
\begin{equation}
 2\alpha_t \left( \dot{x}_t - \frac{\beta_t}{m} \right) = \dot{\beta}_t + V'_t .
 \label{eq18}	
\end{equation}
Now, since $\alpha_t$ is complex and the right-hand side of Eq.~(\ref{eq18}) is real, the
following identities must be satisfied:
\begin{equation}
 \dot{x}_t = \frac{\beta_t}{m} , \qquad \dot{\beta}_t = - V'_t ,
 \label{eq19}
\end{equation}
which is in compliance with Ehrenfest's theorem, i.e., the centroid of the wave packet
evolves in time according to the classical Hamiltonian equations, with $x_t$ and $p_t$
defining a classical trajectory.
Thus, from now on, we will denote $\beta_t$ as $p_t$.
Accordingly, Eq.~(\ref{eq15}) can be recast in a simpler manner, as
\begin{equation}
 \dot{\gamma}_t = \frac{i\hbar\alpha_t}{m} + p_t \dot{x}_t - E_t
   + \frac{\lambda\hbar}{m}\ \alpha_t ,
 \label{eq20}
\end{equation}
where $E_t = p_t^2/2m + V_t$.

Therefore, the set of equations of motion that determine the time evolution of the ansatz
wave function (\ref{eq8}) is
\begin{eqnarray}
 \dot{x}_t & = & \frac{p_t}{m} ,
 \label{eq21} \\
 \dot{p}_t & = & - V'_t ,
 \label{eq22} \\
 \dot{\alpha}^r_t & = & - \frac{2 (\alpha^r_t)^2}{m}
  + (1 - \lambda)\ \frac{2 (\alpha^i_t)^2}{m} - \frac{1}{2} V''_t ,
 \label{eq23} \\
 \dot{\alpha}^i_t & = & - \frac{4 \alpha^r_t \alpha^i_t}{m} ,
 \label{eq24} \\
 \dot{\gamma}^r_t & = & -(1 - \lambda)\ \frac{\hbar\alpha^i_t}{m} + p_t \dot{x}_t - E_t ,
  \label{eq25} \\
 \dot{\gamma}^i_t & = & \frac{\hbar\alpha^r_t}{m} .
 \label{eq26}
\end{eqnarray}
To this set of equations, we need to add the general expression for the equation of motion
of the trajectories, which reads as \cite{sanz-bk-2}
\begin{equation}
 \dot{x} = \frac{1}{m} \left[ p_t + 2 \alpha^r_t (x - x_t) \right] .
 \label{eq27}
\end{equation}
The trajectories obtained from this equation for each case will help us to understand the
role played by the classicality-enforcing potential in terms of the coupling strength $\lambda$.
Note that, if $\lambda = 0$, the above system of equations reproduces the original
Heller's set \cite{sanz-bk-2,heller:JCP:1975}, while for $\lambda = 1$, the real part
of both $\alpha_t$ and $\gamma_t$ decouples from the imaginary part of $\alpha_t$,
which is related to the dispersion of the wave packet.
This is better appreciated if Eq.~(\ref{eq27}) is recast as
\begin{equation}
 \frac{\dot{x} - p_t/m}{x - x_t} = \frac{2\alpha^r_t}{m} ,
 \label{eq27b}
\end{equation}
which can readily be integrated to yield the general expression
\begin{equation}
 x(t) = x_t + \left[ x(0) - x_0 \right] \exp \left[ \int_0^t \frac{2\alpha^r_{t'}}{m}\ dt' \right] .
 \label{eq27bb}
\end{equation}
Accordingly, any effect induced by $\alpha^i_t$ on the Bohmian trajectories disappears,
as Eq.~(\ref{eq27}) will only depend on $\alpha^r_t$.
On the other hand, the phase accumulated by the wave, accounted for by $\gamma^r_t$,
will equal the classical action, $\mathcal{S} = \int \mathcal{L}\ dt$, where
$\mathcal{L} = p_t \dot{x}_t/2 - V_t$ is the classical Lagrangian.

From Eqs.~(\ref{eq23}) and (\ref{eq24}), we note that, in the fully nonlinear regime $\lambda = 1$, the dependence of $\alpha_t^r$ on the imaginary part of $\alpha_t$ is washed out, although $\alpha^i_t$ still depends on $\alpha^r_t$. In general terms, this ``non-recriprocal'' dependence can be understood by noting that $\alpha^r_t$ is directly related to dispersion in the trajectory dynamics, according to Eq.~(\ref{eq27}). In turn, these dynamics must be in compliance with the dispersion undergone by the wave packet (\ref{eq10}), which is described by $\alpha^i_t$. Therefore, for a dispersive wave packet, $\alpha^r_t$ must not only be nonzero, but should also influence the dynamics of $\alpha^i_t$, even if the opposite is not the case, regardless of the value of $\lambda$.

%%%%%%%%%%%%%%%%%%%%%%%%%%%%%%%%%%%%%%%%%%%%%%%%%%%%%%%%%%%%%%%%%%%%%%%%
%%%%%%%%%%%%%%%%%%%%%%%%%%%%%%%%%%%%%%%%%%%%%%%%%%%%%%%%%%%%%%%%%%%%%%%%

\section{Single wave packet dynamics}
\label{sec3}

%%%%%%%%%%%%%%%%%%%%%%%%%%%%%%%%%%%%%%%%%%%%%%%%%%%%%%%%%%%%%%%%%%%%%%%%

\subsection{Free-space propagation}
\label{sec31}

In the case of free-space propagation, $V(x)=0$, the set of equations (\ref{eq21}) to (\ref{eq26})
becomes
\begin{eqnarray}
 \dot{x}_t & = & \frac{p_t}{m} ,
 \label{eq28} \\
 \dot{p}_t & = & 0 ,
 \label{eq29} \\
 \dot{\alpha}^r_t & = & - \frac{2 (\alpha^r_t)^2}{m}
  + (1 - \lambda)\ \frac{2 (\alpha^i_t)^2}{m} ,
 \label{eq30} \\
 \dot{\alpha}^i_t & = & - \frac{4 \alpha^r_t \alpha^i_t}{m} ,
 \label{eq31} \\
 \dot{\gamma}^r_t & = & -(1 - \lambda)\ \frac{\hbar\alpha^i_t}{m} + p_t \dot{x}_t - E_t ,
 \label{eq32} \\
 \dot{\gamma}^i_t & = & \frac{\hbar\alpha^r_t}{m} .
 \label{eq33}
\end{eqnarray}
If the initial ansatz is
\begin{equation}
 \psi(x,0) = \left( \frac{1}{2\pi\sigma_0^2} \right)^{1/4}
  \exp \left[ - \frac{(x-x_0)^2}{4\sigma_0^2} + \frac{i p_0 (x-x_0)}{\hbar} \right] ,
\end{equation}
we have the following initial conditions for the Gaussian parameters:
\begin{equation}
 \alpha_0 = \frac{i\hbar}{4\sigma_0^2} , \qquad
 \gamma_0 = \frac{i\hbar}{4}\ \ln \left( 2\pi\sigma_0^2 \right) ,
 \label{eq35}
\end{equation}
and $(x_0,p_0)$ for $(x_t,p_t)$.
For the linear case, the solutions are readily obtained, being
\begin{eqnarray}
 x_t & = & x_0 + \frac{p_0}{m}\ t, \\
 p_t & = & p_0 , \\
 \alpha_t & = & \frac{i\hbar}{4\sigma_0^2} \left( \frac{1}{1 + i\hbar t/2m\sigma_0^2} \right)
 = \frac{i\hbar}{4\sigma_0 \tilde{\sigma}_t} , \\
 \gamma_t & = & \gamma_0 + \frac{i\hbar}{2}\ \ln \left( 1 + \frac{2 \alpha_0 t}{m} \right)
 = \frac{i\hbar}{4}\ \ln \left( 2\pi\tilde{\sigma}_t \right) ,
\end{eqnarray}
where
\begin{equation}
 \tilde{\sigma}_t = \sigma_0 \left( 1 + \frac{i\hbar t}{2m\sigma_0^2} \right) .
 \label{eq40}
\end{equation}
We thus obtain the usual Gaussian dispersion or diffraction, described by the wave function
\begin{equation}
 \psi(x,t) = \left( \frac{1}{2\pi\tilde{\sigma}_t^2} \right)^{1/4}
  \exp \left[ -\frac{(x-x_0)^2}{4\sigma_0\tilde{\sigma}_t} + \frac{ip_0 (x - x_t)}{\hbar} - \frac{E_t t}{\hbar} \right] ,
 \label{eq41}
\end{equation}
and the typical hyperbolic trajectories describing these dynamics,
\begin{equation}
 x(t) = x_t + \frac{\sigma_t}{\sigma_0} \left[ x(0) - x_0 \right] ,
 \label{eq44}
\end{equation}
where
\begin{equation}
 \sigma_t = |\tilde{\sigma}_t| =
 \sigma_0 \sqrt{ 1 + \left( \frac{\hbar t}{2m\sigma_0^2} \right)^2 } .
 \label{eq43}
\end{equation}

The above results and their dynamical consequences are well-known
(see Refs.~\cite{sanz:JPA:2008,sanz-bk-2} for a general discussion).
In order to better understand the physics associated with the $\alpha_t$-parameter as well
as its contribution to the quantumness of the system, we now consider a more general initial
condition $\alpha_0$, such that $\alpha^r_0 \ne 0$.
In this case, the integration in time of Eqs.~(\ref{eq30}) and (\ref{eq31}) renders
\begin{equation}
 \alpha_t = \frac{\alpha_0}{1 + 2 \alpha_0 t/m} ,
\end{equation}
where its respective real and imaginary parts
\begin{eqnarray}
 \alpha^r_t & = & \frac{\alpha^r_0}{1 + 4\alpha_0^r t/m + (2 |\alpha_0| t/m)^2}
  + \frac{2 |\alpha_0|^2 t/m}{1 + 4\alpha_0^r t/m + (2 |\alpha_0| t/m)^2} ,
  \label{eq45} \\
 \alpha^i_t & = & \frac{\alpha^i_0}{1 + 4\alpha_0^r t/m + (2 |\alpha_0| t/m)^2} .
  \label{eq45imag}
\end{eqnarray}
Substituting Eq.~(\ref{eq45}) into Eq.~(\ref{eq27bb}), and then integrating in time, leads to the trajectory equation
\begin{equation}
 x(t) = x_t + \sqrt{1 + 4\alpha_0^r t/m + (2 |\alpha_0| t/m)^2} \left[ x(0) - x_0 \right] ,
 \label{eq49}
\end{equation}
which, somehow, resembles the functional form displayed by Eq.~(\ref{eq44}), although it now contains an additional linear term, as it also does the denominator of $\alpha^i_t$ (apart from substituting the prefactor $\alpha_0^i$ by $|\alpha_0|$ in the quadratic term).
The presence of this term introduces a linear correction in the quadratic expansion at very short times, i.e., for $t \ll m/2|\alpha_0|$, where Eq.~(\ref{eq49}) can be approximated by
\begin{equation}
 x(t) \approx x(0) + \left\{ \frac{p_0}{m} + \frac{2\alpha_0^r}{m} \left[ x(0) - x_0 \right] \right\} t + \frac{2|\alpha_0|^2}{m^2} \left[ x(0) - x_0 \right] t^2 .
 \label{eq49approx}
\end{equation}
As it can readily be seen, taking $p_0 = 0$ for simplicity, if $t \ll \alpha_0^r/m|\alpha_0^2|^2$, the linear term dominates the dynamics in the very short time regime, unlike the case of a typical minimum uncertainty wave packet seen above.
This effect is purely induced by the addition of an initial extra position-dependent phase, which is the role played by a nonzero $\alpha_0^r$ [note that, in such a case, the wave packet acquires an initial phase given by $\alpha_0^r (x - x_0)$ at $t=0$].
This is the case, for instance, if we wish to induce self-focusing without the intervention of any additional nonlinear term in the Schr\"odinger equation, as it shown in \cite{sanz:arxiv:2023}.

With respect to the other limiting case, namely, the full nonlinear scenario, for $\lambda = 1$, the initial condition (\ref{eq35}) for a minimum uncertainty wave packet leads to
\begin{eqnarray}
 \alpha_t & = & \frac{i\hbar}{4\sigma_0^2} , \\
 \gamma_t & = & \frac{i\hbar}{4}\ \ln \left( 2\pi\sigma_0 \right) ,
\end{eqnarray}
and, therefore, the ``classical'' wave function will read as
\begin{equation}
 \psi(x,t) = \left( \frac{1}{2\pi\sigma_0^2} \right)^{1/4}
  \exp \left[ -\frac{(x-x_t)^2}{4\sigma_0^2} + \frac{i p_0(x - x_t)}{\hbar} - \frac{iEt}{\hbar} \right] .
 \label{eq50}
\end{equation}
This wave function is very similar to the wave function (\ref{eq41}), except for the lack of
a dispersive term, since its width is constant in time, as one might expect once the dispersive
contribution accounted for by $\alpha_t^i$ is suppressed.
This gives rise to the phenomenon of spatial localization, which is precisely the behavior
that one would expect for a free-propagating Gaussian classical statistical particle 
distribution, where all its constituents (an ensemble of identical, non-interacting
particles) have exactly the same momentum $p_0$.
Accordingly, given that the phase factor is linear in the position, the Bohmian trajectories
will be straight lines described by the equation
\begin{equation}
 x(t) = x(0) + \frac{p_0}{m}\ t ,
\end{equation}
as it corresponds to uniform motion.

\begin{figure}[!t]
\centering
\includegraphics[width=\columnwidth]{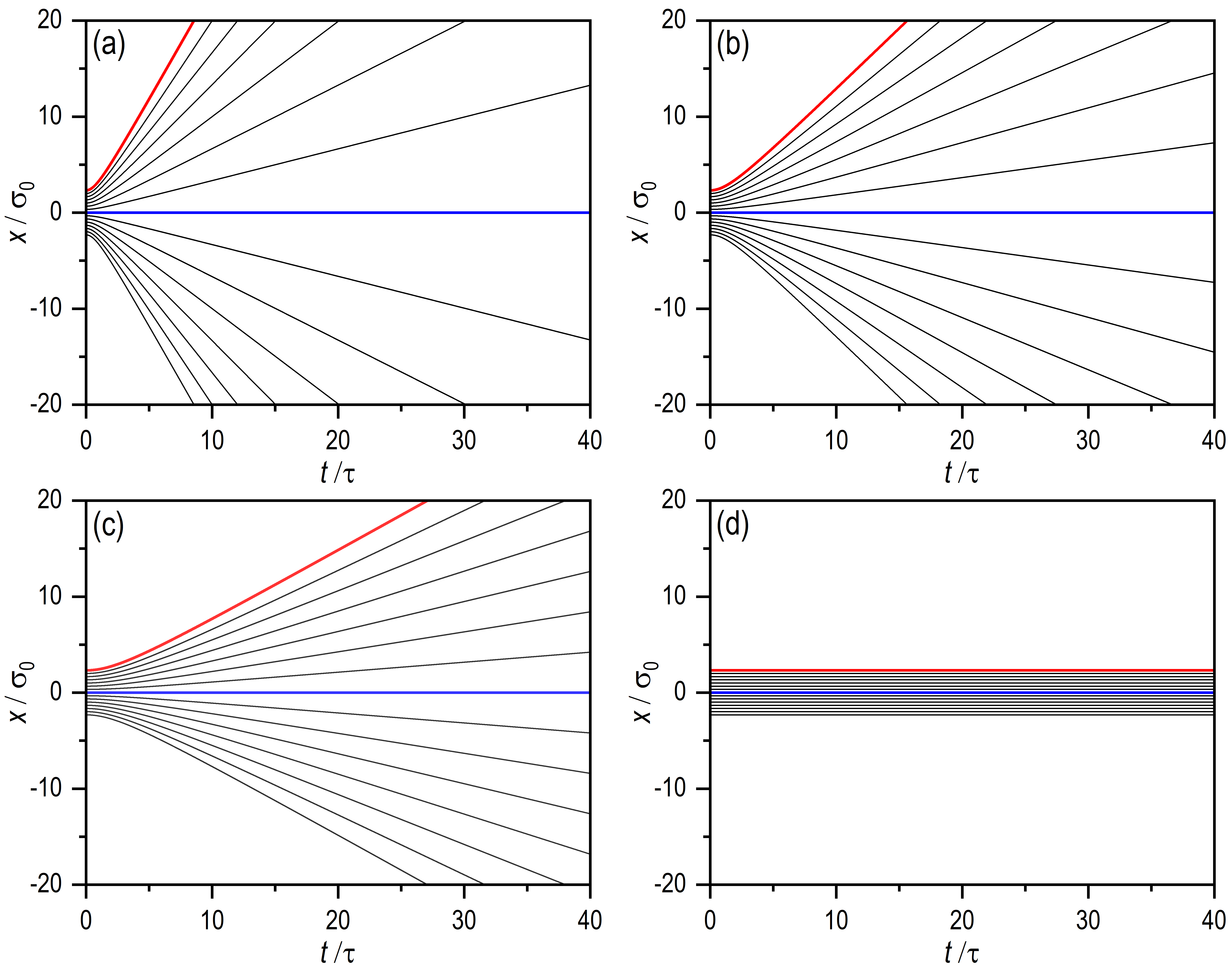}
\caption{\label{fig1}
 Suppression of the dispersion of a single Gaussian wave packet with an initial width
 $\sigma_0 = 0.5$ and increasing coupling constant: (a) $\lambda = 0$, (b) $\lambda = 0.7$,
 (c) $\lambda = 0.9$, and (d) $\lambda = 1$.
 A set of 15 Bohmian trajectories has been used to illustrate the process and, in turn, to
 show how these trajectories eventually show a classical-like behavior, in the Newtonian
 sense,	for $\lambda = 1$ in panel (d).
 The initial conditions for these trajectories are such that the blue trajectory has been
 launched from the center of the wave packet, while the red one starts at a $2.5 \sigma_0$-distance
 from it.
 In all simulations here, we have considered $\hbar = 1$, $m = 1$, $x_0 = 0$, and $p_0 = 0$.}
\end{figure}

As we have seen, the classical wave function (\ref{eq50}) is dispersionless and has a
definite energy, $E$, which translates into Bohmian trajectories all with the same momentum.
This result seems to confirm the hypothesis that the quantum potential is directly responsible for
dispersion or diffraction phenomena, since the rectilinear motion is unaffected, at least whenever $\alpha_0$ is purely imaginary.
The four sets of trajectories plotted in Fig.~\ref{fig1} correspond to four different values
of the strength coupling constant, $\lambda$, ranging from a full quantum regime to a full
classical one: (a) $\lambda = 0$, (b) $\lambda = 0.7$, (c) $\lambda = 0.9$, and
(d) $\lambda = 1$.
These trajectories have been obtained by numerically integrating the set of coupled
equations (\ref{eq28}) to (\ref{eq33}) together with Eq.~(\ref{eq27}).
As it can be seen, the increase of $\lambda$ leads to a cancellation of the diffraction
undergone by the wave packet.
The trajectories represented with blue and red solid lines help to better visualize how,
as $\lambda$ increases, the whole set approaches the dispersionless condition.

Usually, within the full quantum regime, an increase in $\sigma_0$ (or in the mass $m$)
gives rise to slowing down the diffraction process, since the effective width of the
wave packet, described by Eq.~(\ref{eq43}), will remain nearly constant
($\sigma_t \approx \sigma_0$)  for longer times.
The characteristic time that determines when diffractive effects are going to become
apparent~is
\begin{equation}
 \tau = \frac{2m\sigma_0^2}{\hbar} .
 \label{eq52}
\end{equation}
However, the action of a non-vanishing classicality-enforcing potential generates an analogous
inverse effect, as it annihilates the imaginary component in (\ref{eq40}), which is
equivalent to diminishing the relevance of the time scale $\tau$.

From the above results, it is thus clear that the presence of an classicality-enforcing potential
nonlinear contribution in the Schr\"odinger equation leads to motions that, at least in
appearance, are classical-like.
This, however, does not constitute a proof itself of the classicality of the system in the
regime $\lambda = 1$, neither that the dispersion is only attainable to the quantum potential.
This is a misleading impression that arises from the particular type of initial ansatz here considered, with a vanishing $\alpha_0^r$, because we were describing the propagation of a usual minimum uncertainty wave packet.
If, on the contrary, we consider a non-vanishing real part for $\alpha_0$, then the real and imaginary parts for $\alpha_t$ will read as
\begin{eqnarray}
 \alpha^r_t & = & \frac{\alpha^r_0}{1 + 2 \alpha^r_0 t/m} ,
 \label{eq52b} \\
 \alpha^i_t & = & \frac{\alpha^i_0}{\left( 1 + 2\alpha_0^r t/m \right)^2} ,
 \label{eq52c}
\end{eqnarray}
and, from Eq.~(\ref{eq52b}), we obtain the trajectory equation
\begin{equation}
 x(t) = x_t + \left( 1 + \frac{2 \alpha^r_0 t}{m} \right) [x(0) - x_0]
 = x(0) + \left\{ \frac{p_0}{m}
   + \frac{2 \alpha^r_0}{m}\ [x(0) - x_0] \right\} t ,
 \label{eq58tra}
\end{equation}
which indicates that the trajectories will spread with a different rate depending on
their distance with respect to the center of the wave packet, $x_0$, something that also
happens in the fully linear regime ($\lambda = 0$) for large $t$ \cite{sanz:AJP:2012}.
Nonetheless, here we notice that the term proportional to $t^2$ has disappeared from these expressions, so the wave packet will show a linear expansion at any time, according to
Eq.~(\ref{eq52c}), and unlike what happened for $\lambda = 0$, where this linear behavior  was observable only either for a very short time at the beginning of the propagation, or asymptotically in the long-time regime.
This behavior is also made evident by the trajectories rendered by Eq.~(\ref{eq58tra}), which is identical to Eq.~(\ref{eq49approx}), except for the quadratic dependence on time.
However, although we have a linear dependence on time, note that we cannot relate this behavior to a proper classical dynamics, as the addition of a phase affects the propagation of the swarm of particles, which are going to disseminate in different directions depending on their position with respect to the center of the distribution, $x_0$, and also the initial value $\alpha_0^r$.

From the above results, summing up, it is worth noting that, regardless of a full action of the quantum potential, dispersion has to do, in general, with the presence of a non-vanishing real part of $\alpha_t$.
The quantum potential contributes, as we have seen, to generate with time the development of a non-vanishing $\alpha_t^r$ if $\alpha_0$ is a pure imaginary quantity.
In such a case, the addition of the classicality-enforcing potential leads to the cancellation of the real part of $\alpha_t$ generated by the quantum potential.
However, if $\alpha_0^r$ is nonzero, not only such a cancellation does not occur for $\lambda = 1$, but the wave packet still undergoes a linear dispersion since the very beginning, which somehow mimics a diffusion problem.

Apart from the ``non-classical'' behavior that we have just found for a non-vanishing
$\alpha^r_t$, an important distinctive trait of classical trajectories is that they can get
across the same spatial point at the same instant, because of the multi-valuedness of the
momentum.
In the present context, therefore, this means that a genuine classical behavior should lead
to a violation of the well-known Bohmian non-crossing rule \cite{sanz:JPA:2008}.
To investigate this hypothesis, next we consider two scenarios that should help to prove it
or disprove it, namely, the harmonic oscillator and the two wave-packet interference.

%%%%%%%%%%%%%%%%%%%%%%%%%%%%%%%%%%%%%%%%%%%%%%%%%%%%%%%%%%%%%%%%%%%%%%%%

\subsection{Harmonic oscillator}
\label{sec32}

Consider a Gaussian wave packet acted by a harmonic potential,
\begin{equation}
	V(x) = \frac{1}{2}\ m \omega^2 x^2 .
 \label{eq54a}
\end{equation}
In the linear case ($\lambda = 0$), the general solution to Heller's equations (\ref{eq21}) 
to (\ref{eq26}) is
\begin{eqnarray}
  x_t & = & x_0 \cos \omega t + \frac{p_0}{m\omega}\ \sin \omega t ,
  \label{eq54} \\
%    = X_0 \sin (\omega t - \varphi_0) , \\
  p_t & = & p_0 \cos \omega t - m\omega x_0 \sin \omega t ,
  \label{eq55} \\
%    = -m\omega X_0 \sing (\omega t - \varphi_0) , \\
 \alpha_t & = & \frac{m\omega}{2} \left(
   \frac{ 2\alpha_0/m\omega - \tan \omega t}{1 + (2\alpha_0/m\omega) \tan \omega t} \right)
 = \frac{m\omega}{2} \left(
   \frac{ 2\alpha_0 \cos \omega t - m\omega \sin \omega t}{2\alpha_0 \sin \omega t + m\omega \cos \omega t} \right) ,
  \label{eq56} \\
  \gamma_t & = & - \frac{i\hbar}{4}\ \ln \left( \frac{2\alpha^i_0}{\pi\hbar} \right)
  + \frac{i\hbar}{2}\ \ln \left(\cos \omega t + \frac{2\alpha_0}{m\omega} \sin\omega t\right)
 \nonumber \\
 & & + \frac{1}{2\omega} \left( \frac{p_0^2}{2m}
     - \frac{1}{2}\ m\omega^2 x_0^2 \right) \sin 2\omega t - p_0 x_0 \sin^2 \omega t .
 \label{eq57}
\end{eqnarray}
%
%where $X_0 = \sqrt{x_0^2 + (p_0/m\omega)^2}$ and $\varphi_0 = \tan^{-1} (p_0/m\omega x_0)$.
These solutions describe the dynamics of a ``breathing'' Gaussian wave packet, i.e., a wave
packet with an oscillatory width as it propagates back and forth between the turning points
of the potential function (\ref{eq54a}), $x_\pm = \pm \sqrt{2E/m\omega^2}$.

The general expression for the real and imaginary parts of $\alpha_t$ read as
\begin{eqnarray}
 \alpha_t^r & = & \frac{m\omega}{2} \left( \frac{ - \sin \omega t \cos \omega t + (2\alpha_0^r/m\omega) \sin 2 \omega + (2|\alpha_0|/m\omega)^2 \cos \omega t \sin \omega t}{\cos^2 \omega t + (2\alpha_0^r/m\omega) \sin 2 \omega t + (2|\alpha_0|/m\omega)^2 \sin^2 \omega t} \right) ,
 \label{eqnew1} \\
 \alpha_t^i & = & \frac{\alpha_0^i}{\cos^2 \omega t + (2\alpha_0^r/m\omega) \sin 2 \omega t + (2|\alpha_0|/m\omega)^2 \sin^2 \omega t} ,
 \label{eqnew2}
\end{eqnarray}
from which we readily obtain the general expression for the corresponding Bohmian trajectories,
\begin{equation}
 x(t) = x_t + \left[ \cos^2 \omega t + \left(\frac{2\alpha^r_0}{m\omega}\right) \sin 2 \omega t + \left(\frac{2|\alpha_0|}{m\omega}\right)^2 \sin^2 \omega t \right]^{1/2} \left[ x(0) - x_0 \right] .
\end{equation}
If now we choose the particular initial condition $\alpha_0 = im \omega/2\hbar$, from Eq.~(\ref{eq56}) we readily obtained that $\alpha_t = \alpha_i$, that is, the width of the wave packet is constant all the way down during its back and forth journey, with its wave function being
\begin{eqnarray}
 \psi(x,t) & = & \left( \frac{m\omega}{\pi\hbar} \right)^{1/4}
  \exp \left[ - \frac{m\omega}{2\hbar}\ (x - x_t)^2 + \frac{ip_t x}{\hbar} - i \left( \frac{p_0^2}{4m\omega\hbar} - \frac{m\omega x_0^2}{4\hbar} \right) \sin 2\omega t \right.
 \nonumber \\ & & \qquad \qquad \qquad \left. - \frac{i p_0 x_0}{\hbar}\ \cos^2 \omega t - \frac{i\omega t}{2} \right] ,
%  \exp \left[ - \frac{m\omega}{2\hbar}\ (x - x_t)^2 + \frac{ip_t x}{\hbar} - \frac{i (p_0^2/2m - m\omega^2 x_0^2/2) \sin 2\omega t}{2\hbar\omega} - \frac{i p_0 x_0 \cos^2 \omega t}{\hbar} - \frac{i\omega t}{2} \right] ,
 \label{eq62}
\end{eqnarray}
which has an invariant probability density profile, because its width remains constant at any time, $\sigma_0 = \sqrt{\hbar/2m\omega}$, although it moves back and forth inside the harmonic well.
Correspondingly, its space-dependent phase factor is linear with the $x$-coordinate,
which gives rise to the trajectory equation
\begin{equation}
 x(t) = x_t + \cos \omega t \left[ x(0) - x_0 \right] ,
\end{equation}
which indicates that all associated Bohmian trajectories are parallel one another and also
with respect to the classical one, described by $x_t$.
The Gaussian wave packet in this specific case is called a Glauber coherent state, which is regarded
as the most classical wave function that we may have, because, as seen above (and made more evident
by the Bohmian trajectories), it reproduces the behavior of a classical oscillator.
For any other arbitrary value of $\alpha_0$, but still with a zero real part ($\alpha_0^r=0$), we have the so-called squeezed coherent states, for which $\alpha_t$ oscillates every half a period between $\alpha_0$ (at $\omega t = 0, \pi/2, \pi, \ldots$) and $-m^2\omega^2/4\alpha_0$ (at $\omega t = \pi/4, 3\pi/4, \ldots$), thus generating the ``breathing'' type propagation mentioned above.
In this case, the Bohmian trajectories follow the equation
\begin{equation}
 x(t) = x_t + \left[ \cos^2 \omega t
 + \left(\frac{2\alpha^i_0}{m\omega}\right)^2 \sin^2 \omega t \right]^{1/2} \left[ x(0) - x_0 \right] ,
\end{equation}
where the time-dependent prefactor in the second term accounts for the ``breathing'' of the wave packet.

For simplicity in our analysis, we are going to consider Glauber states to remove any
additional misleading element in the dynamics (e.g., ``breathing'').
Note that, since these states are regarded as the most classical ones, their associated
trajectories constitute a very valuable tool to probe their dynamics, and hence to detect
any particularity in the trend towards the classical limit.
Thus, proceeding as in the previous section, after integrating Eq.~(\ref{eq23}) for
$\lambda = 1$, we obtain
\begin{equation}
 \alpha^r_t = \frac{m\omega}{2} \frac{(2\alpha^r_0/m\omega) - \tan \omega t}{1 + (2\alpha^r_0/m\omega) \tan \omega t} ,
 \label{eq65}
\end{equation}
which is independent of the imaginary part of $\alpha_t$, although $\alpha^i_t$ still depends on $\alpha^r_0$ [see Eq.~(\ref{eq65b}) below].
If $\alpha^r_t$ is now substituted into Eq.~(\ref{eq27}) and then we integrate in time,
we readily obtain
\begin{equation}
 x(t) = x_t + \left\arrowvert \cos \omega t + \frac{2\alpha^r_0}{m\omega}\ \sin \omega t \right\arrowvert \left[ x(0) - x_0 \right] .
\end{equation}
In the case of the coherent state, where $\alpha_t = im\omega/2$ ($\alpha^r_0 = 0$), this
solution simplifies to
\begin{equation}
	x(t) = x_t + |\cos \omega t | \left[ x(0) - x_0 \right] .
 \label{eq67}
\end{equation}
which resembles the classical solution, except for the fact that Bohmian trajectories are
kept either on one side or the other of the central one, described by the classical
trajectory $x_t$.
That is, they are prevented from coming together on the same point in the strict limit
$\lambda = 1$, where such a point refers to the spatial loci where a homologous set of
classical trajectories focus on during the course of a full oscillation.

\begin{figure}[!t]
	\centering
	\includegraphics[width=\columnwidth]{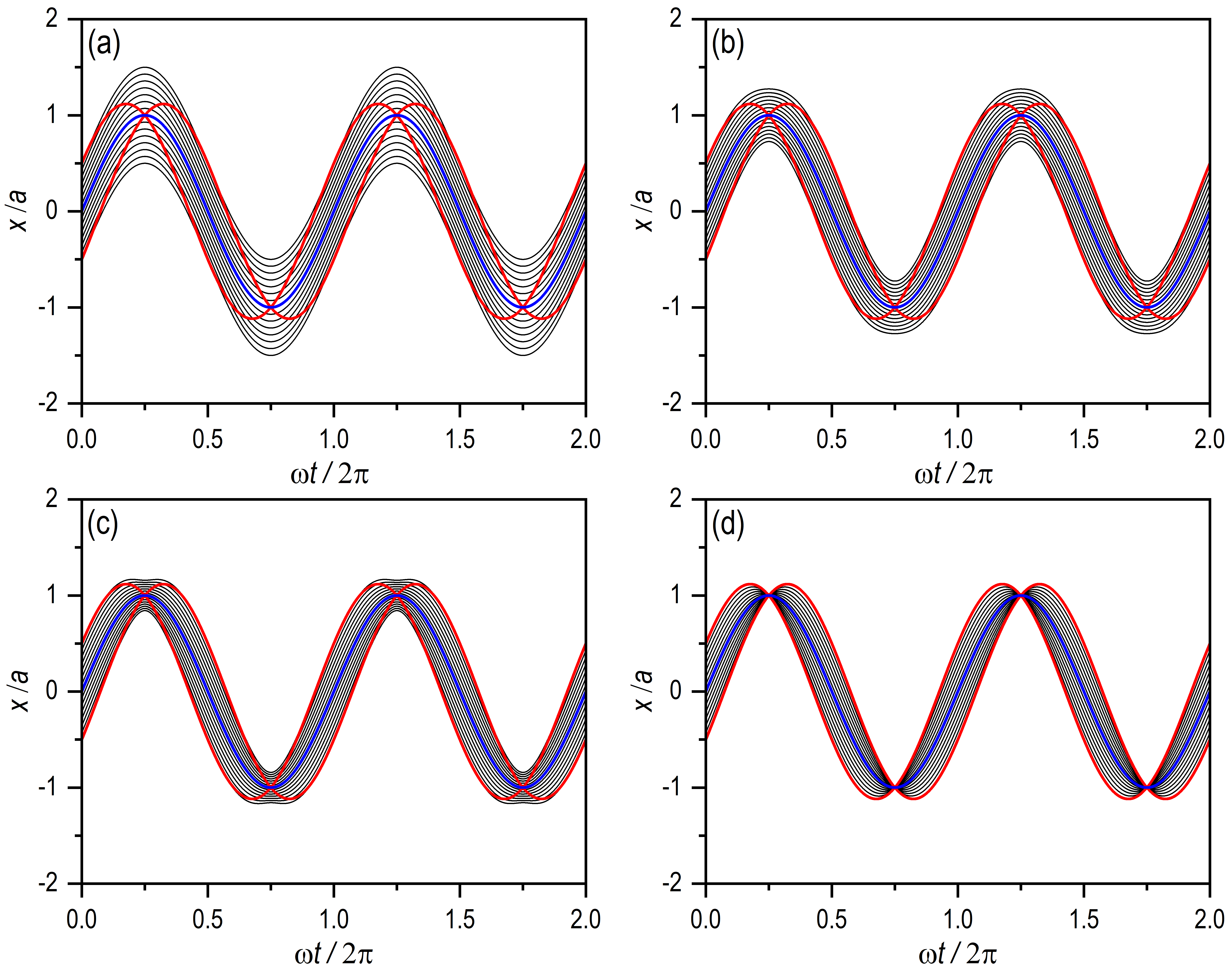}
	\caption{\label{fig2}
		Gradual approach to a classical-like behavior, in a Newtonian sense, for a coherent wave
		packet in a harmonic oscillator with $\omega = 1$ and launched from the center of the
		potential, at $x_0 = 0$, with $p_0 = 1$, as the coupling constant increases:
		(a) $\lambda = 0$, (b) $\lambda = 0.7$, (c) $\lambda = 0.9$, and (d) $\lambda = 0.999$.
		A set of 15 Bohmian trajectories has been used to illustrate the process, noticing that, as
		$\lambda$ becomes close to the unity, there is a kink denoting the position where the
		corresponding classical trajectories merge.
		Three classical trajectories superimposed to the set of Bohmian trajectories to show the
		turning points.
		The initial conditions for these trajectories are such that
		the blue trajectory has been launched from the center of the wave packet, while the two
		red ones start with the same initial conditions of their Bohmian counterparts at the
		margins.
		In all simulations here, we have considered $\hbar = 1$ and $m = 1$.}
\end{figure}

It is worth highlighting that the sets of trajectories described by Eq.~(\ref{eq67}) are in compliance with the time-dependence acquired now by the width of the wave packet, unlike the fully linear case, where it is constant ($\sigma_0 = \sqrt{\hbar/2m\omega}$, as indicated above). This time-dependence can readily be obtained after substituting Eq.~(\ref{eq65}) into Eq.~(\ref{eq24}), and then solve for $\alpha_t^i$, which renders
\begin{equation}
 \alpha^i_t = \frac{\alpha_0^i}{\left[ \cos \omega t + (2\alpha^r_0/m\omega) \sin \omega t \right]^2} .
 \label{eq65b}
\end{equation}
Defining, as in previous cases, that the width at any later time is $\sigma_t = \sqrt{i\hbar/4\alpha_t^i}$, we obtain
\begin{equation}
 \sigma_t = \frac{i\hbar}{4\alpha_0^i} \left\arrowvert \cos \omega t + \frac{2\alpha_0^r}{m\omega}\ \sin \omega t \right\arrowvert .
\end{equation}
This is general expression for the width of the ``classical'' wave packet inside the well, which describes a time-periodic variation for the width, analogous to a squeezed state.
Now, if we apply the condition here that the wave packet is a coherent Glauber state, with $\alpha_0^r = 0$, then the above expression reads as
\begin{equation}
 \sigma_t = \sigma_0 |\cos \omega t | ,
\end{equation}
i.e., the time-dependence of the width of the wave packet is consistent with the motion displayed by the trajectories, periodically turning from a maximum value (given by
$\sigma_0$) to extreme focusing (zero width), and back again.
As it can be seen, this is independent of the initial conditions considered, so we will observe maximum width at even multiples $\omega t/4\pi$, and focusing at odd multiples of this quantity.

\begin{figure}[!t]
	\centering
	\includegraphics[width=\columnwidth]{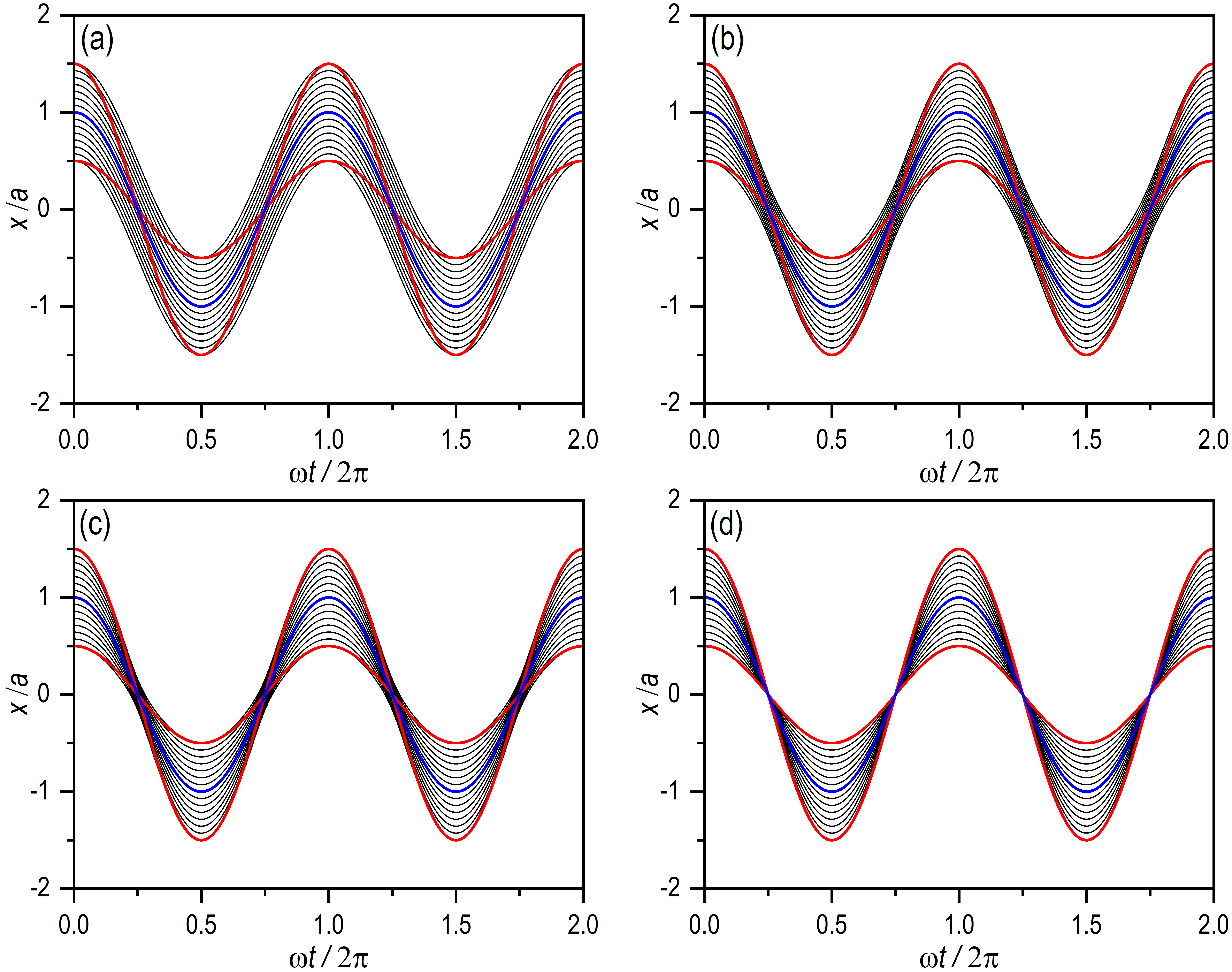}
	\caption{\label{fig3}
		Gradual approach to a classical-like behavior, in a Newtonian sense, for a coherent wave
		packet in a harmonic oscillator with $\omega = 1$ and launched from the turning point
        $x_0 = 1$, with $p_0 = 0$, as the coupling constant increases:
		(a) $\lambda = 0$, (b) $\lambda = 0.7$, (c) $\lambda = 0.9$, and (d) $\lambda = 0.999$.
		A set of 15 Bohmian trajectories has been used to illustrate the process, noticing that, as
		$\lambda$ becomes close to the unity, there is a kink denoting the position where the
		corresponding classical trajectories merge.
		Three classical trajectories superimposed to the set of Bohmian trajectories to show the
		turning points.
        The initial conditions for these trajectories are such that
		the blue trajectory has been launched from the center of the wave packet, while the two
		red ones start with the same initial conditions of their Bohmian counterparts at the
		margins.
		In all simulations here, we have considered $\hbar = 1$ and $m = 1$.}
\end{figure}

To illustrate the transition towards the classical limit, in Figs.~\ref{fig2} and
\ref{fig3} we have plotted sets of trajectories associated with wave packets launched from
$x_0 = 0$ with $p_0 = \sqrt{2mE}$, and from $x_0 = x_+ = \sqrt{2E/m\omega^2}$ and $p_0 = 0$,
respectively.
In both cases, in order to make more apparent the loci where classical trajectories focus,
we have included classical trajectories launched from the center of the initial wave packet
(denoted with blue solid line) and also the same initial conditions as the two Bohmian
trajectories launched from the margins (denoted with red solid line).
Also, for an easier comparison, in both cases we have used the same four values for the
strength coupling constant: (a) $\lambda = 0$, (b) $\lambda = 0.7$, (c) $\lambda = 0.9$,
and (d) $\lambda = 0.999$.
In Fig.~\ref{fig2}, we observe a gradual transition from the characteristic parallel motion
displayed by Bohmian trajectories in a fully linear regime ($\lambda = 0$) to the incipient
appearance of focal points in the turning points, $x_\pm$, which are the spatial loci where
the classical trajectories with the same initial conditions meet together.
Indeed, in the limit $\lambda \to 1$, in Fig.~\ref{fig2}(d), the Bohmian and the classical
trajectories are seemingly the same.
The same behavior is observed in Fig.~\ref{fig3}, although the focal point now appear at the
center of the potential function, $x=0$, as it is clearly shown by the crossing undergone
by the classical trajectories at that point.
As it is clearly seen, the set of Bohmian trajectories gradually focuses on those points as
$\lambda$ increases, until it is difficult to distinguish any difference between Bohmian
and classical trajectories started at the same initial position.

\begin{figure}[!t]
	\centering
	\includegraphics[width=\columnwidth]{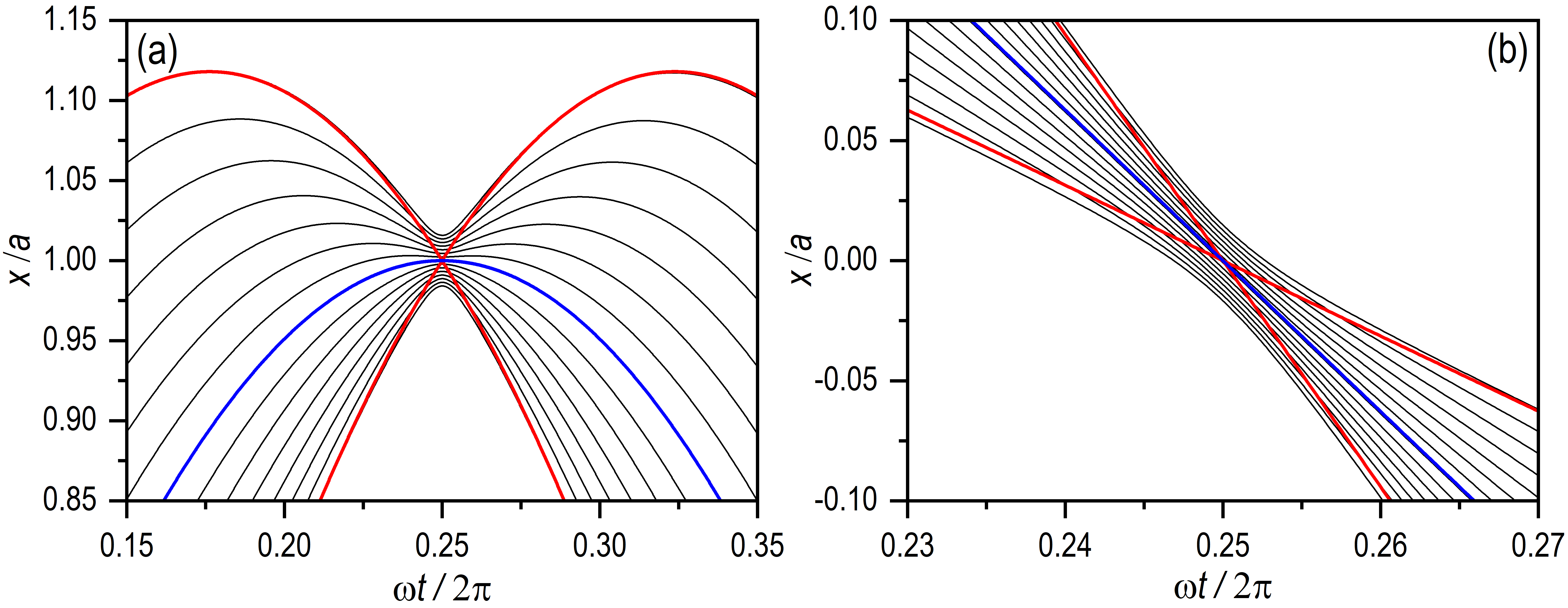}
	\caption{\label{fig4}
      To better appreciate the focusing region in Figs.~\ref{fig2} and \ref{fig3}, here an enlargement
      of these figures is shown in panels~(a) and (b), respectively, around the respective first focal
      points, at $\omega t = 0.25$, for $\lambda = 0.999$.
      As in the previous figures, regarding the classical trajectories at display, the blue one has been
      launched from the center of the wave packet, while the two red trajectories start with the same
      initial conditions of their Bohmian counterparts at the margins.
      In all simulations here, we have considered $\hbar = 1$ and $m = 1$.}
\end{figure}

A closer look at the foci in both cases shows, as it is seen in Fig.~\ref{fig4}, that the
approach is fast but leaving the Bohmian trajectories on one side or the other with respect
to the central one (blue solid lines), unlike the behavior displayed by the classical
trajectories (red solid lines), which cross at the foci.
Although the simulations have been carried for $\lambda = 0.999$, the trend is the same as
we approach more and more the limit $\lambda \to 1$.
Therefore, this behavior disproves the fact that including an classicality-enforcing potential leads
to the classical limit.
There are important differences, as we have seen in the case of the Glauber state,
which is regarded as the most classical quantum state.

%%%%%%%%%%%%%%%%%%%%%%%%%%%%%%%%%%%%%%%%%%%%%%%%%%%%%%%%%%%%%%%%%%%%%%%%
%%%%%%%%%%%%%%%%%%%%%%%%%%%%%%%%%%%%%%%%%%%%%%%%%%%%%%%%%%%%%%%%%%%%%%%%

\section{Two wave-packet interference dynamics}
\label{sec4}

%%%%%%%%%%%%%%%%%%%%%%%%%%%%%%%%%%%%%%%%%%%%%%%%%%%%%%%%%%%%%%%%%%%%%%%%

\subsection{Free-space propagation}
\label{sec41}

\begin{figure}[!t]
\centering
\includegraphics[width=\columnwidth]{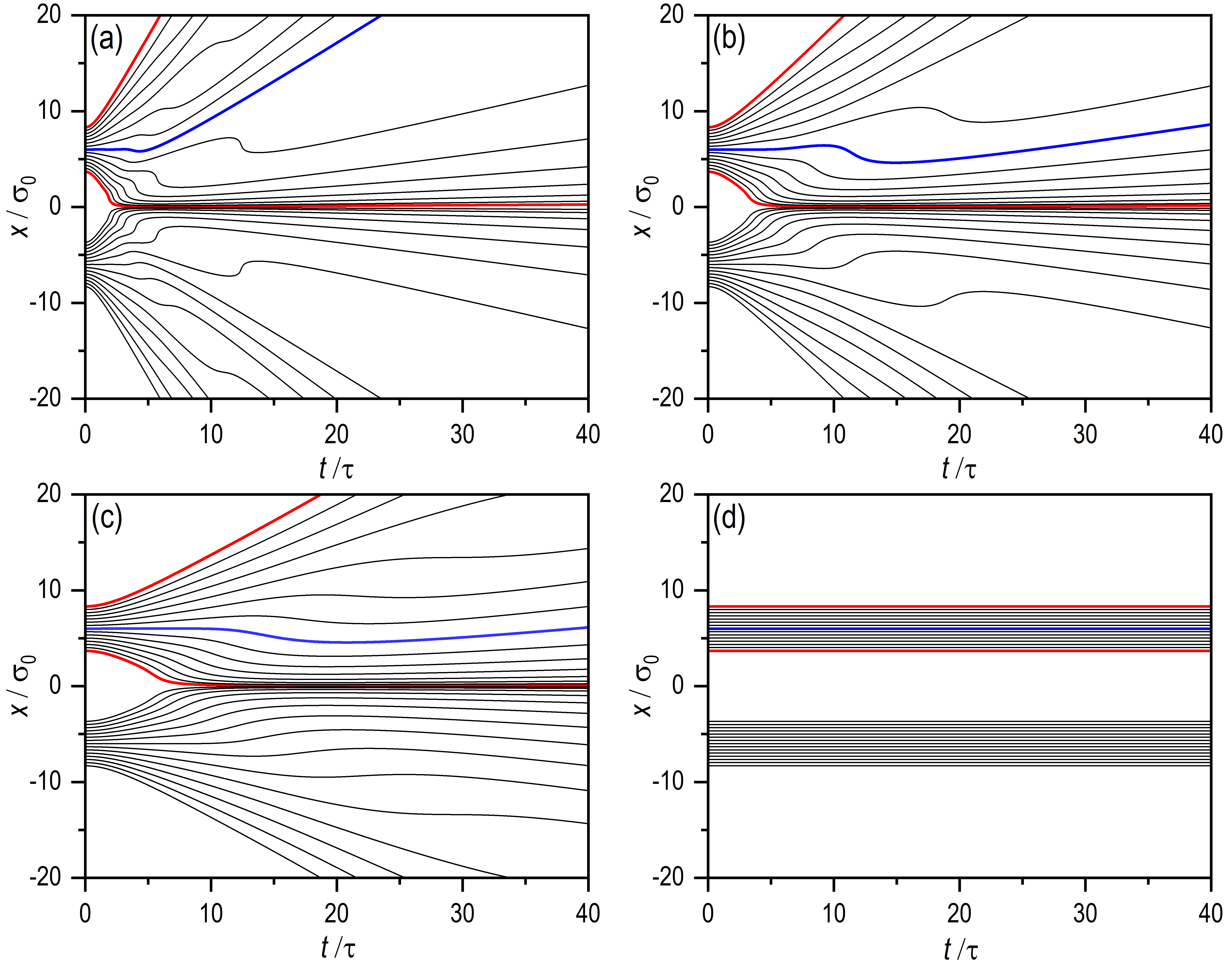}
\caption{\label{fig5}
 Suppression of the dispersion of two Gaussian wave packets in a coherent superposition,
 both with an initial width $\sigma_0 = 0.5$ and centered at $x_\pm = \pm x_0$ ($x_0 = 3$),
 as the coupling constant increases: (a) $\lambda = 0$, (b) $\lambda = 0.7$, (c) $\lambda = 0.9$,
 and (d) $\lambda = 1$.
 A set of 15 Bohmian trajectories has been launched from region covered by each wave packet
 in order to illustrate the process and, in turn, to show how these trajectories eventually
 show a classical-like behavior, in the Newtonian sense, for $\lambda = 1$, as seen in
 panel (d).
 As it can be noticed, the lack of dispersion also produces that the wave packet cannot
 overlap and hence the trajectories will not accumulate along the directions expected from a
 typical interference process.
 The initial conditions for these trajectories are such that
 the blue trajectory has been launched from the center of the wave packet starting on the
 positive part of the $x$-axis, while the two red ones start at a $2.5 \sigma_0$-distance
 from the previous one on either side.
 In all simulations here, we have considered $\hbar = 1$, $m = 1$, and $p_0 = 0$.}
\end{figure}

Apart from using the harmonic oscillator to investigate the classical limit, as seen in
Sec.~\ref{sec32}, one still may wonder whether there are other typical quantum traits that
cannot be fully suppressed by this classicality-enforcing potential, such as interference.
To investigate this scenario, let us remember that, in free space, we can have two
interference situations worth discussing separately, namely, Young-type interference or
interference produced by a head-on collision of two wave packets \cite{sanz:JPA:2008}.
In both cases, we start with a superposition of two identical wave packets, like
(\ref{eq50}), localized at two different positions and with different momenta,
\begin{eqnarray}
 \psi(x,t) & = & \sqrt{N} \left( \frac{1}{2\pi\sigma_0^2} \right)^{1/4}
  \left\{ \exp \left[ -\frac{(x-x_{1,t})^2}{4\sigma_0^2} + \frac{i p_{1,0}(x - x_{1,t})}{\hbar} - \frac{iE_1 t}{\hbar} \right] \right.
 \nonumber \\ & & \qquad \qquad \qquad \qquad
  \left. + \exp \left[ -\frac{(x-x_{2,t})^2}{4\sigma_0^2} + \frac{i p_{2,0}(x - x_{2,t})}{\hbar} - \frac{iE_2 t}{\hbar} \right] \right\} .
 \label{eq68}
\end{eqnarray}
Here, $N$ is the normalization factor of the superposition; if the center-to-center distance between the
wave packets is such that $|x_{1,0} - x_{2,0}| \gg \sigma_0$, then $N \approx 1/\sqrt{2}$.
To simplify, we consider $x_{1,0} = - x_{2,0} = x_0$ and $p_{1,0} = - p_{2,0} = - p_0$.

\begin{figure}[!t]
\centering
\includegraphics[width=\columnwidth]{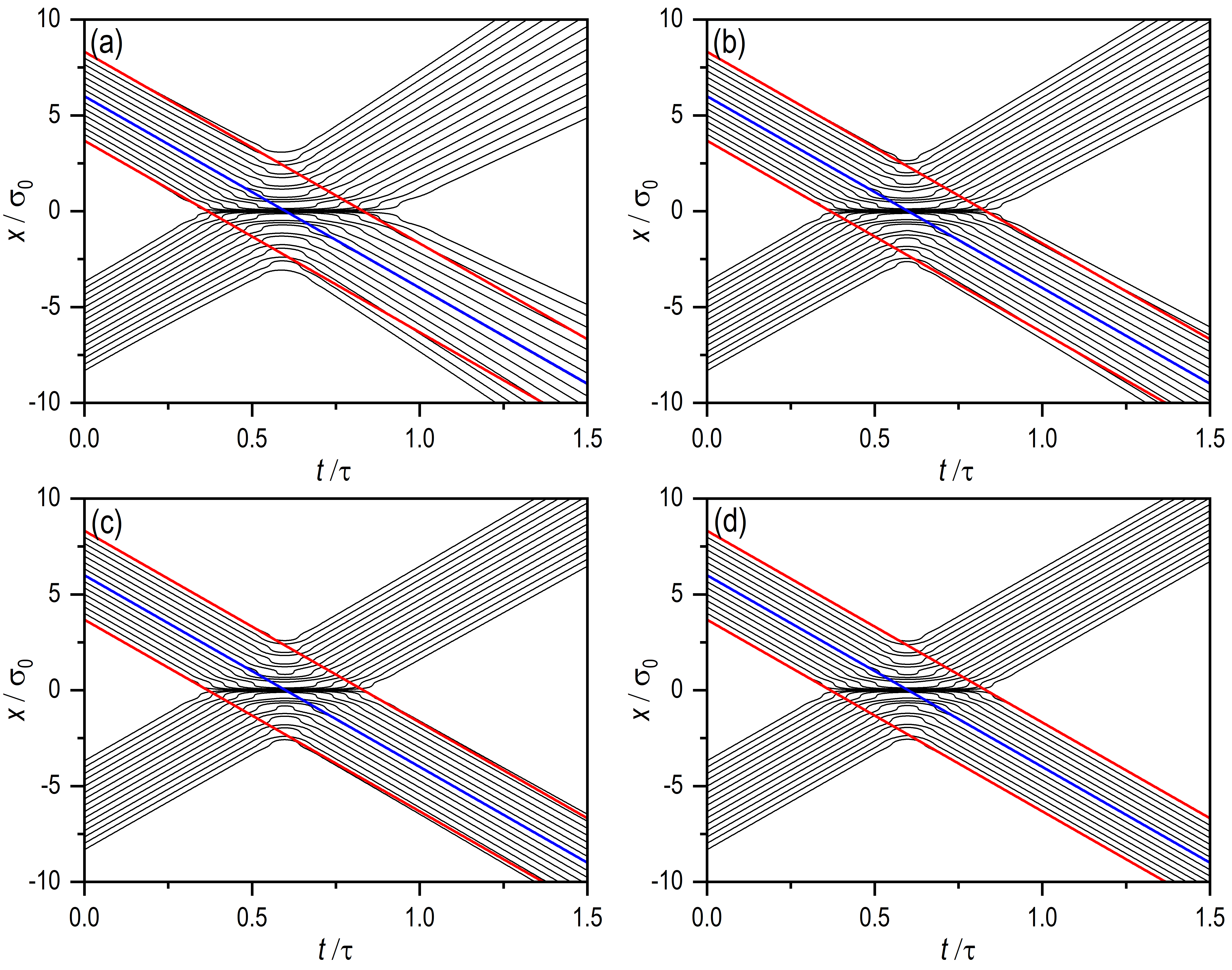}
\caption{\label{fig6}
 Same as Fig.~\ref{fig3}, but adding an extra transverse momentum to each wave packet
 ($p_\pm = \mp 10$), so that they fully overlap at $t = 3$.
 In this case, we find that the Bohmian trajectories do not cross the center of symmetry of
 the figure, although true classical trajectories would be able to do it, as it is inferred
 from the behavior displayed by the three classical trajectories included in the graphs.
 The initial conditions for these trajectories are such that the blue trajectory has been
 launched from the center of the wave packet starting on the positive part of the $x$-axis,
 while the two red ones start at a $2.5 \sigma_0$-distance from the previous one on either side.
 In all simulations here, we have considered $\hbar = 1$ and $m = 1$.}
\end{figure}

As it is explained and discussed with detail in \cite{sanz:JPA:2008}, a Young-type situation
arises if $p_0 = 0$, because the gradual spreading of the wave packets eventually provokes
their spatial overlapping and, hence, the appearance of interference fringes.
Thus, in the long time limit, the interference profile displays a stationary shape, where
the distance between consecutive minima (or maxima) increases linearly with time
\cite{sanz:ENTROPY:2023}.
From a trajectory point of view, this phenomenon consists of two distinctive stages.
In a first stage, the trajectories associated with each partial wave behave, in a good
approximation, in an independent manner, showing the typical hyperbolic dispersion already
seen in Sec.~\ref{sec31} for a free Gaussian wave packet.
Once the two waves start overlapping importantly, this behavior is severely distorted,
transitioning towards the second stage, where all the trajectories, as a single ensemble,
start redistributing in space along different channels, each one directly related to an
interference maximum.
This is the behavior observed in Fig.~\ref{fig5}(a), where we have put with different
color several trajectories launched from one of the initial wave packets (with blue the
one launched from the center, and with red those launched from the margins).
As $\lambda$ increases, because the dispersion of the two initial wave packets is inhibited,
the appearance of the channel sets of trajectories undergoes a longer and longer delay.
In the limit $\lambda = 1$, in Fig.~\ref{fig5}(d), because the dispersion of the two wave
packets is totally inhibited, we simply observe two separated sets of parallel trajectories.

Again, similarly to the case of a single Gaussian, if we only examine the Young-type
scenario, one could conclude that all signatures of quantumness have been removed.
However, if we consider the head-on collision of two wave packets, characterized by
two wave packets that do not almost increase their width during the course of the
propagation, but that show interference when they meet at an intermediate position,
the situation is very different.
This scenario is illustrated in Fig.~\ref{fig6}.
As $\lambda$ increases, we notice that, effectively, the dispersion of the wave packets
is inhibited; in the limit $\lambda = 1$, indeed, the Bohmian trajectories coincide with
the classical ones at the margins (red solid lines).
However, in analogy to the behaviors already observed with the harmonic oscillator, the
divergence from a true classical behavior immediately arises if we observe what happens
in the crossing region, which here is much larger than in the case of the oscillator.
First, as $\lambda$ increases, interference features within this region do not disappear
at all, but persist even for $\lambda = 1$.
Second, while the classical trajectories get across this interference region, the
homologous pairs of Bohmian trajectories undergo a deflection backwards, such that the
output part of a classical trajectory overlaps with the output part of a Bohmian trajectory
corresponding to the opposite wave packet.
Therefore, here we find another important behavior, which revels that the presence of the
classicality-enforcing potential does not prevents the observation of typically quantum features
(in this case, interference and non-crossing).

%%%%%%%%%%%%%%%%%%%%%%%%%%%%%%%%%%%%%%%%%%%%%%%%%%%%%%%%%%%%%%%%%%%%%%%%

\begin{figure}[!t]
	\centering
	\includegraphics[width=\columnwidth]{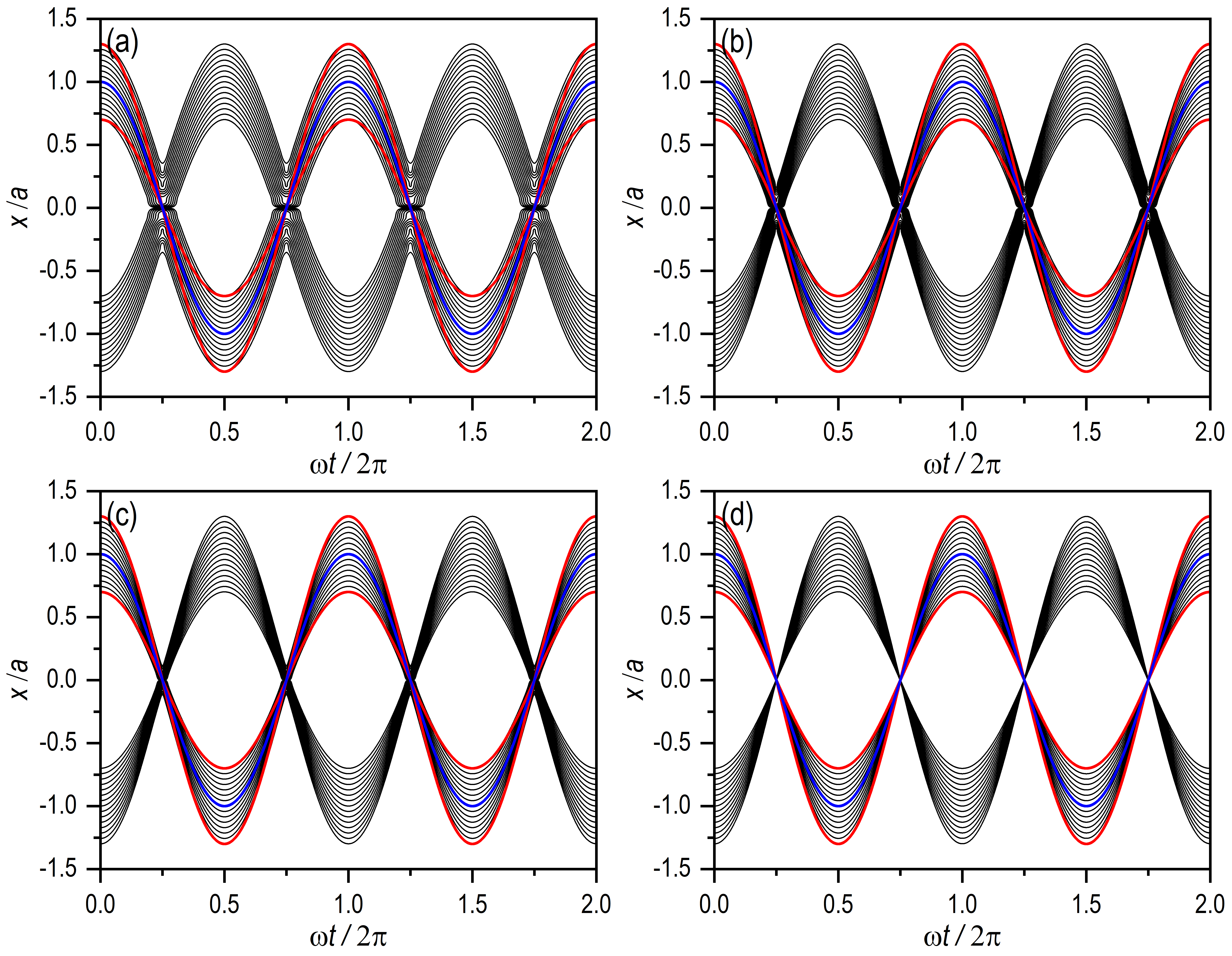}
	\caption{\label{fig7}
		Same as Fig.~\ref{fig3}, but for two wave packets inside the harmonic oscillator, each
		starting on an opposite turning point at $x_\pm = \pm x_0$ ($x_0 = 5$): (a) $\lambda = 0$,
		(b) $\lambda = 0.7$, (c) $\lambda = 0.9$, and (d) $\lambda = 0.999$.
		In this case, we find that the Bohmian trajectories do not cross the center of symmetry of
		the figure as $\lambda$ increases, although true classical trajectories would be able to do
		it, as the three classical trajectories superimposed make apparent.
		The initial conditions for these trajectories are such that
		the blue trajectory has been launched from the center of the wave packet, while the two
		red ones start with the same initial conditions of their Bohmian counterparts at the
		margins.
		In all simulations here, we have considered $\hbar = 1$ and $m = 1$.}
\end{figure}

\subsection{Harmonic oscillator}
\label{sec42}

Last, we consider another interference process, but this time inside a harmonic well, with
two counter-propagating wave packets launched from opposite turning points, $x_\pm$.
The superposition state is as described by (\ref{eq68}), substituting the free-propagating
Gaussian wave packets, but the wave packets for a harmonic oscillator (\ref{eq62}).
The results for these superpositions are represented in Fig.~\ref{fig7}, where we have
considered again four different values for the coupling strength constant: (a) $\lambda = 0$,
(b) $\lambda = 0.7$, (c) $\lambda = 0.9$, and (d) $\lambda = 0.999$.
In Fig.~\ref{fig7}(a), we observe the appearance of a series of voids at the region where
the two wave packets overlap.
These voids arise as a consequence of the presence of nodes in the wave function, where the
phase is undefined and hence also the velocity field, thus preventing the trajectories from
crossing these regions and making them to undergo sudden turns, as we can also see in
Fig.~\ref{fig6}(a).
Now, because Bohmian trajectories cannot cross, they undergo a bounce backwards, reaching
again their initial positions in half a cycle, unlike the corresponding classical
trajectories (denoted with blue and red solid lines), which require a full cycle.
As $\lambda$ increases, though, the Bohmian trajectories seem to mimic the behavior observed
in the classical trajectories, as it is seen in Fig.~\ref{fig7}(d), with the ensembles coming
from each wave packet focusing at the center of the potential function.

\begin{figure}[!t]
	\centering
	\includegraphics[width=\columnwidth]{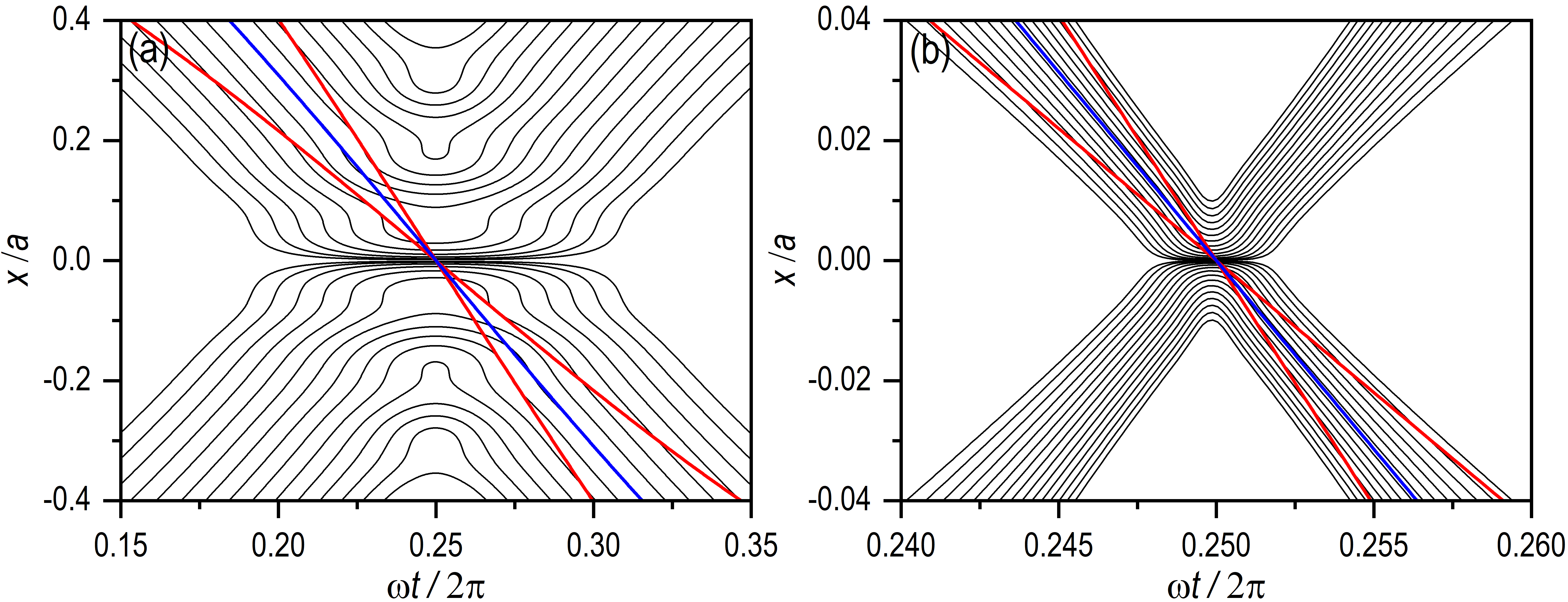}
	\caption{\label{fig8}
		To better appreciate the details of the region where the two sets of Bohmian trajectories
		undergo the bounce backwards in Figs.~\ref{fig7}a ($\lambda = 0$) and \ref{fig7}d
		($\lambda = 0.999$), panels (a) and (b) show here, respectively, an enlargement around
		of the corresponding regions.
		The maximum approach takes place at $\omega t = 0.25$.
		As in Fig.~\ref{fig7}, the initial conditions for the trajectories plotted are such that
		the blue trajectory has been launched from the center of the wave packet, while the two
		red ones start with the same initial conditions of their Bohmian counterparts at the
		margins.
		In all simulations here, we have considered $\hbar = 1$ and $m = 1$.}
\end{figure}

To better understand what happens at the focus, in Fig.~\ref{fig8} we show an enlargement
in a neighboring region around it for the cases displayed in Figs.~\ref{fig7}(a) and \ref{fig7}(b),
where we can clearly observe the presence of voids in the first case, and the focusing effect in
the second one.
Two remarks are worth noting.
First, note the order of magnitude of difference between the two regions displayed despite
in both cases we have used exactly the same initial conditions, which gives an idea of the
induced focusing.
Should we have used a $\lambda$-value closer to 1, the dimensions of the focus region would
have decreased even more.
Second, although we do not observe the presence of nodes in Fig.~\ref{fig8}(b), this does not
mean that they do not form.
In symmetric head-on collisions (i.e., whenever both wave packets have exactly the same
properties and they propagate with the same speed in opposite directions), the nodes formed
when both wave packets fully overlap are separated by a distance $\Delta x = \pi\hbar/p_0$
provided the dispersion at that time with respect is negligible compared to the initial
dispersion.
In all panels in Fig.~\ref{fig6}, this is precisely the case, so all the nodes are separated
a distance $\Delta x \approx 0.31$ (their positions can be inferred from the sudden turns
undergone by the trajectories at $t=0.3$).
In the case of the two wave packet inside the harmonic well, the distance between nodes
is given by the expression $\Delta x = \pi\hbar/m\omega x_0 \approx 0.63$, as it
can be clearly seen in Fig.~\ref{fig8}(a).
As the focusing becomes stronger for sets of trajectories with the same initial conditions,
it is clear that they will pack closer and closer, in distances of orders much below
$\Delta x$, and hence we will not be able to see them, which is the case of
Fig.~\ref{fig8}(b).
To do so, and hence to obtain a representation analogous to that of Fig.~\ref{fig6}(d),
we would therefore need to consider a much wider range for initial conditions around the
center of each wave packet.
Of course, this would imply considering to cover with initial conditions regions far away
from the centroids of the wave packets, where the probability density already is already
negligible (note that in all cases here we have chosen initial conditions reaching regions
with very low values of the probability density).

%%%%%%%%%%%%%%%%%%%%%%%%%%%%%%%%%%%%%%%%%%%%%%%%%%%%%%%%%%%%%%%%%%%%%

\section{Final remarks}
\label{sec5}

Here we have analyzed the consequences of introducing nonlinearities in the Schr\"odinger
equation in the form of the so-called Bohm's quantum potential.
In the literature, this has been widely regarded as a means to recover the classical limit,
since, following the usual Bohmian prescription \cite{holland-bk}, the addition of this
classicality-enforcing potential gives rise to an equation of motion for the phase field identical
to the classical Hamilton-Jacobi equation.
Hence, if from this latter equation one obtains the trajectories of classical mechanics,
from the phase of the classical wave function resulting from the corresponding
nonlinear Schr\"odinger equation one should also obtain Bohmian trajectories that behave
like classical trajectories.
Now, this hypothesis omits the fact that, while in classical mechanics it is allowed a multiple valuation of the momentum (also given by the gradient of the classical action), the
Schr\"odinger equation is constructed under the implicit assumption that their solutions
are single-valued, except for a constant global phase factor.
This means that the phase of the wave function is also single valued, except for that
constant factor, but the velocity field has to be single-valued.
The most striking consequence from this constraint is the well-known Bohmian non-crossing
rule.
Nonetheless, within a more general perspective, this is nothing but a direct physical
consequence arising from the preservation of the phase coherence, present in any wave
theories, regardless of the physical system described and, in particular, of whether
the corresponding equation is linear or nonlinear.

To analyze the above mentioned consequences led by the classicality-enforcing potential, we have
considered the propagation of Gaussian wave packets both in free space and inside of a
harmonic potential, as well as the corresponding superpositions.
Given that Gaussian wave packets are exact solutions of the Schr\"odinger equation for
potential functions that are polynomials of up to second order, and that the addition of
the classicality-enforcing potential is analogous to partly considering a kinetic-type contribution,
we have obtained exact solutions in all cases for both the linear and the fully nonlinear
cases.
For any other intermediate case, we have obtained analytical equations of motion for the
parameters describing the evolution of the Gaussian as well as for the corresponding
trajectories.
Numerically integrating the latter, we have investigated the behavior for various values
of the coupling strength constant, from $\lambda = 0$ to $\lambda = 1$.

From the results obtained, we conclude that the action of the classicality-enforcing potential,
although provides us with behaviors analogous to those displayed by classical trajectories,
it also reveals that its presence does not prevent the observation of typically
quantum features directly connected with the preservation of the quantum coherence, such
as interference and non-crossing.
This situation is similar to that found when dealing with trajectories that describe fully
incoherent quantum systems, but that are obtained from the associated reduced density
matrix \cite{sanz:EPJD:2007}.
In these situations, groups of trajectories associated with two different wave packets do
not cross the symmetry axis between them, because the phase coherence connected to the
information about having the two wave packets involved at the same time in the system
dynamics is still preserved (even though any interference trait is fully washed out).
Obtaining a genuine classical behavior thus requires a suppression of the wave packet that
the trajectories are not associated with \cite{sanz:CPL:2009-2}, which is the case when
an environment is present \cite{sanz:ENTROPY:2023}.

%%%%%%%%%%%%%%%%%%%%%%%%%%%%%%%%%%%%%%%%%%%%%%%%%%%%%%%%%%%%%%%%%%%%%

\section*{Acknowledgments}
Financial support from the Spanish Agencia Estatal de Investigaci\'on (AEI) and the European
Regional Development Fund (ERDF) (Grant No. PID2021-127781NB-I00) is
acknowledged.



  \bibliographystyle{plain} 
 % \bibliographystyle{elsarticle-num} 
 % \bibliographystyle{elsarticle-num-names} 
 %  \bibliography{references}

%\end{document}


\end{document}